%% file: dynamical_heating.tex
\documentclass[fleqn,usenatbib]{mnras}

\usepackage{newtxtext,newtxmath}

\usepackage[T1]{fontenc}

\DeclareRobustCommand{\VAN}[3]{#2}
\let\VANthebibliography\thebibliography
\def\thebibliography{\DeclareRobustCommand{\VAN}[3]{##3}\VANthebibliography}

\usepackage{graphicx}	
\usepackage{amsmath}	

\input definitions

\title[Accretion-induced orbital tightening]{Dynamical heating of newborn stars driven by accretion-induced orbital tightening}

\author[Camacho et al. ]{Vianey Camacho$^{1,2}$,  Andrea Bonilla-Barroso$^1$, Javier Ballesteros-Paredes$^1$\thanks{E-mail: j.ballesteros@irya.unam.mx}, 
\newauthor Manuel Zamora-Avilés$^3$, and Luis Aguilar$^4$
\\
\\
$^{1}$Universidad Nacional Aut\'onoma de M\'exico, Instituto de Radioastronom\'ia y Astrof\'isica.\\
Antigua Carretera a P\'atzcuaro 8701, Ex-Hda. San Jos\'e de la Huerta, 58089, Morelia, Michoac\'an, M\'exico
\\
$^{2}$Center of Astronomy and Gravitation, Department of Earth Sciences, National Taiwan Normal University, 88, Sec. 4, Ting-Chou Rd., \\
Wenshan District, Taipei 116, Taiwan R.O.C
\\
$^{3}$Instituto Nacional de Astrof\'isica, \'Optica y Electr\'onica, Luis E. Erro 1, 72840 Tonantzintla, Puebla, M\'exico
\\
$^4$Universidad Nacional Aut{\'o}noma de M{\'e}xico, Instituto de Astronom\'ia, Unidad Acad{\'e}mica en Ensenada\\ Ensenada 22860 BC, M{\'e}xico
}

\date{Accepted XXX. Received YYY; in original form ZZZ}

\pubyear{\the\year{}}

\begin{document}
\label{firstpage}
\pagerange{\pageref{firstpage}--\pageref{lastpage}}
\maketitle

\begin{abstract}
In previous works, we have shown that stars in the Orion and the Lagoon Nebula Clusters, and simulations of collapsing clouds, exhibit constant velocity dispersion as a function of mass, a result described by Lynden-Bell 50 years ago as an effect of a violent relaxation mechanism. In contrast, numerical simulations of turbulent clouds show that newborn massive stars experience stronger dynamical heating than low-mass stars. We analyzed turbulent numerical simulations and found that this effect arises from the fact that, in clouds that are globally turbulence-supported against collapse, massive stars are formed within more massive and denser clumps and in more crowded environments compared to low-mass stars. This allows them to accrete more mass and interact with other stars simultaneously. As they become more massive, their orbits tighten, increasing their velocity dispersion. In contrast, low-mass stars are formed in the periphery of such cores, more separated, and at lower densities. Thus, their velocity dispersion remains lower because they do not accrete as vigorously as massive stars and tend to be more isolated. We call this mechanism ``accretion-induced orbital tightening.'' Our results and previous findings about violent relaxation provide a key observational diagnostic of how to distinguish the dynamic state of star-forming molecular clouds through the kinematics of their newborn stars. 
\end{abstract}




\section{Introduction}\label{sec:intro}

Although it is clear that stars are formed by the collapse of their cold and dense parent molecular cloud, the detailed process in which this occurs is one of the fundamental open problems in astrophysics. One of the main difficulties is that the actual dynamic state of the parent molecular clouds is unclear. For instance, some groups argue that star formation occurs in hierarchically and chaotically collapsing clouds, in a process that lasts a few initial free-fall times \citep[see, e.g.,][and references therein]{Vazquez-Semadeni+19}. In contrast, other groups argue that molecular clouds are supersonically turbulent, and thus, star formation occurs at a low pace over many free-fall times \citep[e.g.,][]{Krumholz+19, Evans+21}. 

There are important reasons to believe that turbulence may play a role in the dynamics of the ISM. For example, the large Reynolds numbers of the interstellar medium in general, and of molecular clouds in particular, \citep[see, e.g.,][]{Elmegreen_Scalo04}, the scaling velocity dispersion-size relation \citep{Larson81}, the fractal appearance of clouds \citep{Falgarone+91}, the large variety of energy injection mechanisms at different scales (jets, winds, supernova explosions, spiral arm shocks \citep{Norman_Ferrara96}), and more importantly, the low star formation rates compared to what they would be if all molecular gas were converted into stars within a typical free-fall time \citep[see, e.g.,][]{Evans+21, Evans+22}. 
However, the modelling of molecular clouds undergoing collapse suggests that although clouds must be turbulent because they do have large Reynolds numbers, much of the observed non-thermal motions must be produced by infalling motions into different centers of collapse \citep[e.g.,][]{Ballesteros-Paredes+99b, Hartmann+01, Vazquez-Semadeni+07, Heitsch_Hartmann08, Heitsch+09, Ballesteros-Paredes+11a, Ballesteros-Paredes+18}. Under this scheme, the star formation rate can be instantaneously large, but stellar feedback from winds, radiation, and supernova explosions breaks the star formation process, returning much of the mass of the cold molecular clouds into the warm, diffuse ISM \citep[e.g.,][]{Colin+13, Grudic+21, Kim+21}.

At first glance, detecting infall motions in molecular clouds may be the best way to disentangle whether the clouds are collapsing. Nevertheless, molecular clouds' highly irregular and fractal nature makes it nearly impossible to prove this on the scale of entire clouds. Thus, an alternative way to understand whether star-forming clouds are in a state of collapse or supported by turbulence is to examine the kinematics of their newborn young stars. 

In previous contributions \citep{Bonilla-Barroso+22, Bonilla-Barroso+24}, we have shown that collapsing clouds in the Solar Neighborhood form clusters in which their stars exhibit constant velocity dispersion, regardless of the bin mass of their stars. This is the natural outcome of violent relaxation, where the fluctuating gravitational potential is the main factor responsible for driving the motions of the stars \citep{Lynden-Bell67}. In contrast, numerical simulations of turbulent molecular clouds, where the gravitational potential evolves slowly, were found to form clusters in which their massive stars become dynamically hotter than their low-mass stars \citep{Bonilla-Barroso+22}.
The tentative explanation for this behaviour provided in the latter work was that, in the simple picture of turbulent clouds, stronger shocks form denser cores and less intense shocks form less-dense cores \citep[e.g.,][]{Padoan_Nordlund02, MacLow_Klessen04, Ballesteros-Paredes+07}. If so, one could expect the core-to-core velocity dispersion to be larger for massive clumps than for low-mass cores. Since high-mass stars are supposed to be formed in massive cores \citep{McKee_Tan03}, and if the stars are born with the velocity of their parent core, one should expect massive stars to have larger velocity dispersion than low-mass stars. 

Nonetheless, massive stars can develop velocity dispersion larger than those of low-mass stars by other means. One possibility is $n$-body collisional relaxation. The other is accretion-induced orbital tightening of pairs and multiple stellar systems. In both cases, the fundamental idea is that since the cloud is turbulence-supported for several free-fall times, $n-$body interactions occur. In the first case, it is well known that massive stars tend to give kinetic energy to low-mass stars during stellar encounters due to the tendency towards energy equipartition of dynamical systems. As a result, massive stars decrease their velocity and sink into the potential well of the cluster, while low-mass stars increase their velocities and move to larger orbits. However, high-mass stars become more bound at the cluster's center, increasing their velocity dispersion, while the less-bounded lower-mass stars tend to slow down as they have to climb up the potential well. The final effect is, thus, the opposite of energy equipartition: massive stars end up with larger velocity dispersions than low-mass stars. 
An extreme case of this situation is the so-called \citet{Spitzer69} instability, where the gravitational potential at the center of the cluster is dominated only by the massive stars, which evolve on their own dynamical timescale \citep[e.g.,][]{Spera+16}. In this case, it is not strange to find that the system also ejects some massive stars \citep[e.g.,][]{Parker+16}. 

\begin{figure*}
    \includegraphics[width=0.94\linewidth]{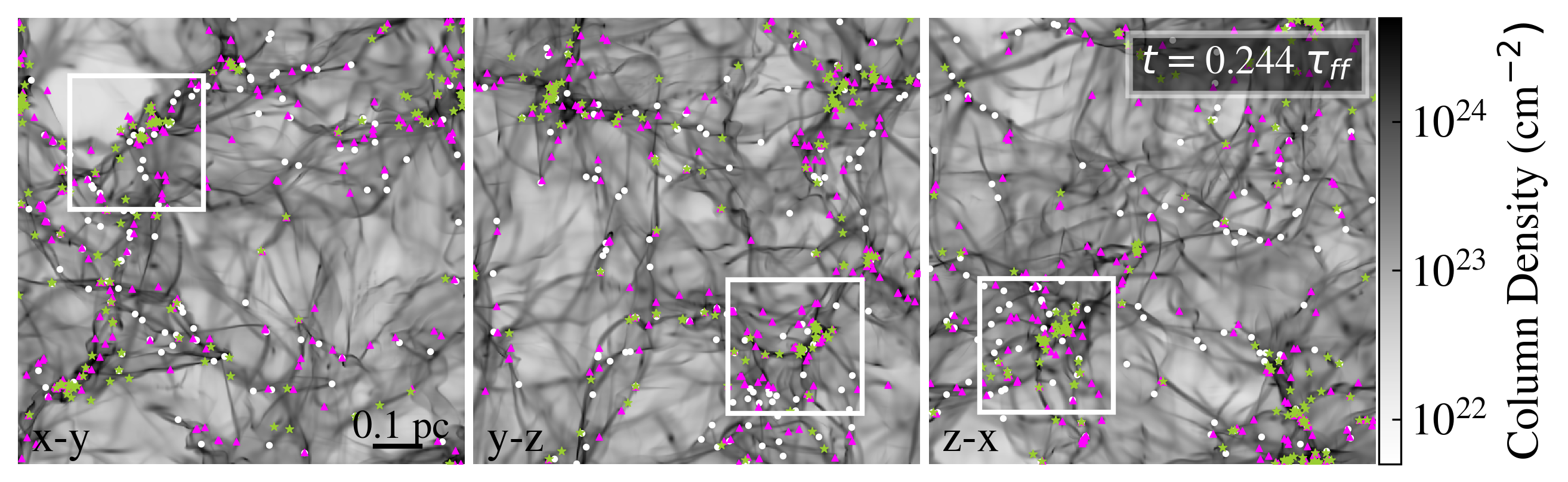}
    \caption{$x-y$, $y-z$ and $z-x$ column density maps of the simulation M10S1. The white dots represent the stellar particles with $M < 1$\Msun, the magenta triangles show the particles with $1\leq M < 5$ \Msun, and the green stars particles with $M \geq 5$\Msun. The smaller box denotes the locus of group~1, also called M10S1-G1, which will be subject of a more detailed analysis.}
    \label{fig:coldens_s1}
\end{figure*}

For the accretion-induced orbital tightening of pairs and multiple stellar systems case, since massive stars tend to be born in denser regions (both in gas and stellar content) than low-mass stars, they will tend to form pairs and/or groups, interacting with each other, tightening their orbits and increasing their velocity dispersion while they become more massive. In contrast, low-mass stars may be born more isolated in lower-density regions. Thus, even if they are in groups, they will not accrete as much mass as the stars in the densest regions of the clouds, and their orbits will not be tightened as high-mass stars, keeping their velocity dispersion lower than massive stars.

As can be seen, in both collisional relaxation, as well as in accretion-induced orbital tightening (or accretion-tightening for short), $n$-body interactions play a pivotal role. However, accretion-tightening specifically necessitates a substantial increase in stellar masses during $n$-body interactions, such that the system's dynamics are significantly modified by the accretion process.
\\ 

In the present work, we analyze numerical simulations of turbulent molecular clouds aimed at understanding how high- and low-mass newborn stars acquire their velocity dispersion during the first epochs of cluster formation and whether high-mass stars are dynamically hotter than low-mass stars due to the turbulent origin of their respective cores, due to $n$-body interactions producing dynamical heating, or due to accretion-induced orbital tightening.
In section \ref{sec:simulations}, we present the numerical simulations, which we analyze in detail in \S\ref{sec:results}. In \S\ref{sec:discussion}, we discuss the three scenarios for the dynamical heating of massive stars and the implications for the models of the dynamical state of molecular clouds. Finally, in \S\ref{sec:conclusions}, we provide our main conclusions.

\section{Numerical simulations}\label{sec:simulations}

Aimed at understanding the primary physical cause of massive stars exhibiting larger velocity dispersion than low-mass stars in simulations of turbulent clouds \citep{Bonilla-Barroso+22, Bonilla-Barroso+24}, we performed numerical MHD simulations of isothermal, self-gravitating molecular clouds with driven turbulence. Our simulations are performed with the FLASH code \citep{Fryxell+00}, and we adopt a scaling such that the numerical box size is 1~pc per side that contains 1000~\Msun.

The box is discretized on a regular $512^{3}$ grid, so the spatial resolution is $\Delta x \simeq 0.0019$~pc $\simeq 391$~AUs. The injected kinetic energy is a mixture of 50$\%$ solenoidal and 50$\%$\ compressive supersonic turbulence, with a power spectrum of $E(k) \propto k^{-2}$, where $k$ is the wavenumber. The turbulence is injected at larger scales, in wavenumbers in the 1$-$3 range, and tuned to achieve a constant Mach number of 10.
To verify our results are statistically robust, we made three different realizations of the same numerical setup but with different random seeds,  i.e., changing the initial turbulent velocity field but keeping the same physics. These simulations will be referred to as M10S1, M10S2, and M10S3.
The gas in our simulations is continuously forced without self-gravity during $\sim$5 crossing times to reach spatial homogeneity statistically. After this time, self-gravity is turned on, and thus, we allow for sink particle formation in order to mimic star formation. During the subsequent evolution, the cloud is sustained on large scales due to the imposed turbulence while it undergoes local collapses. 

The simulations are studied from $t~\sim~0.06~\tauff$ until $\sim$0.25$~\tauff$ (with $\tauff ~=~(3\pi/32G\rho)^{1/2}$ the free-fall time of the whole simulation, and $\rho$ the mean density of the box). The initial time is selected as close as possible to the activation of gravity, ensuring a sufficient number of particles for robust analysis. The final time is determined by the increasing computational cost, as $n$-body interactions significantly slow down the simulations, making further calculations prohibitively expensive in terms of CPU time.

\subsection{Sink particles treatment}

We used the sink particle method implemented by \citet{Federrath+10b} in the FLASH code to create sinks or stellar particles. This method allows gas mass accretion into the particles while computing their dynamical evolution. In brief, this prescription enables the formation of stellar particles within cells at the highest level of reﬁnement that overcome a threshold number density of $n_{\rm th} \simeq 5.5 \times 10^7$cm$^{-3}$ (equivalent to $\rho_{\rm th} \simeq 2.1 \times 10^{-16}$~g~cm$^{-3}$). Around this cell with density $\rho > \rho_{\rm th}$, additional checks are performed in the gas within a spherical control volume of radius $r_{\rm acc}=2.5 \Delta x$: {\it i)} the gas must be converging, {\it ii)} it must be bound (its total energy must be negative), {\it iii)} the gravitational potential must have a local minimum at the central cell, and {\it iv)} the central cell is not within a distance of $r_{\rm acc}$ from another sink. If the last check fails, the excess of mass in the cell ($\Delta m = (\rho - \rho_{\rm th}) \Delta x^3$) is accreted into the existing particle if the cell is gravitationally bound to the sink and if the radial velocity component points towards it. Once formed, the sink particles can move in the cartesian computational domain \citep[see ][for a detailed explanation of the implementation]{Federrath+10b}.

In terms of the gravitational interaction between the gas in the grid and the sink particles the following gravitational forces are calculated: $i)$~gas-gas, to compute the self-gravity of the gas, {using an OctTree Barns-Hut algorithm \citep[the implementation we use is described in ][]{Wunsch+18};} $ii)$~gas-sinks, to compute the gravitational acceleration for the sinks particle due to the gas component, using a ﬁrst-order cloud-in-cell method; $iii)$~sinks-gas, to compute the gravitational acceleration exerted by the sink particles onto the gas cells by a direct sum over all sink particles, and $iv)$ sinks-sinks through a direct summation involving every other sink particle.  For the last two types of interactions, a spline softening is used considering a softening radius equal to the size of the accretion radius, i.e., $r_{\rm soft}=2.5\Delta x$. 
It is important to mention that, in the case of close encounters between sink particles, where the characteristic timestep is significantly shorter than the MHD timestep, the code performs a subcycling of the $n-$body system until the sink system's timestep equals the MHD timestep. 

In order to verify the precision of the $n$-body implementation, \citet{Federrath+10b} performed a series of tests. These authors showed that the algorithm accurately handles close interactions at scales comparable to the grid resolution. In particular, the code can follow stable circular and elliptical orbits, and the accretion of sink particles in a collapsing core (with an initial Bonnor-Ebert or singular isothermal profile) is consistent with analytical estimations. Furthermore, the sink formation and accretion algorithm is in good agreement with other SPH codes \citep[see ][for details]{Federrath+10b}.

\section{Results}\label{sec:results}

Before we proceed, three important notes are in order. First, throughout this work, we indistinctly use the terms `stars' and `sinks'  to refer to our sink particles. However, it should be noted that our sinks do not have stellar feedback, which could modify the kinematics of the newborn stellar groups by modifying the gravitational potential \citep[see, e.g., ][]{Zamora-Aviles+18}. A study addressing these effects is left for a further contribution. Second, although we will focus mainly on the results of Run~{M10S1}, the results we obtain in other realizations with different random and Mach numbers are statistically similar. Thus, we consider our findings are robust. Finally, along this work, we divide our sink particles into three groups: low-mass stars are those sinks with masses $M\leq1~M_\odot$; intermediate-mass stars, with $1< M\leq5~M_\odot$; and high-mass stars, with $M>5~M_\odot$. 
However, to highlight important kinematic differences developed by the stars as they become more or less massive, we also split the stars according to their {\it final} mass. Thus, stars may also be labelled according to whether their mass is larger or lower than the median mass of the studied group of stars. 

In Fig.~\ref{fig:coldens_s1}, we show the column density map of the M10S1 simulation in the $x-y$, $y-z$, and $z-x$ planes at $t=0.244~\tauff$, in grayscale. The different symbols in this figure represent the stellar particles: white circles,  magenta triangles, and green stars denote sinks in the low- intermediate- and high-mass ranges, as defined above. It can be seen that the three projections in this figure show a highly filamentary structure, typical of simulations of turbulence-dominated molecular clouds. The smaller box in each panel denotes a region where the concentration of stars is slightly larger. For further reference, we will call it Group~1 or M10S1-G1. 

\begin{figure}
    \includegraphics[width=0.9\linewidth]{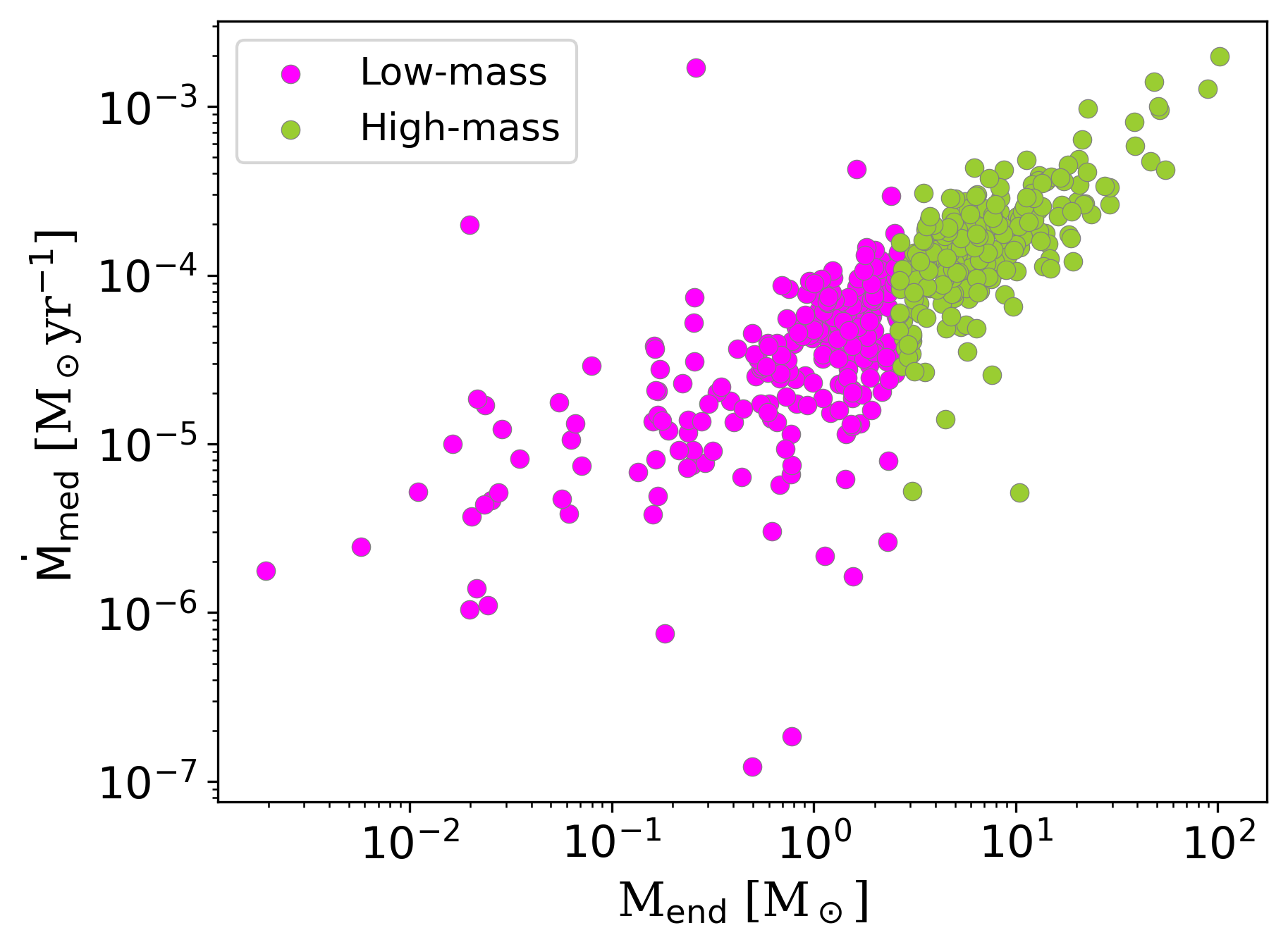}
    \caption{Median accretion rate of the stellar particles in the M10S1 run, computed since the appearance of each sink until the end of the simulation, as a function of their final mass, $M_{\rm end}$. High- (low-) mass stars are defined as sink particles with masses higher (lower) than the median mass of all sinks at the end of the simulation, $M_{\rm med, end} = 2.6$~\Msun. As it can be seen, stars that end up with higher masses have, statistically speaking, larger accretion rates.}
   \label{fig:accrete}
\end{figure}

As mentioned in the previous section, the sink particles in our simulations are allowed to accrete mass, and high-mass stars have, statistically speaking, larger mass accretion rates than the low-mass sinks \citep[see also][and references therein]{Bonnell+07, Ballesteros-Paredes+15}. To show that this is the case, in Fig.~\ref{fig:accrete} we plot the median mass accretion rate $\dot{\mathrm{M}}_{\mathrm{med}}$ of each sink particle in the simulation (computed as the median accretion rate of each sink particle since it was formed until the end of the simulation), as a function of their final mass, $M_{\rm end}$.
In this figure, high- and low-mass stars are classified according to whether their mass is larger or smaller than the median mass 
$M_{\rm med,end}= 2.6 ~M_\odot$ of all sinks at the end of the simulation. We can notice that statistically speaking, the accretion rates of the sinks that end up being massive (green) are larger than those of the sinks that end up being low-mass particles (magenta).
%

\begin{figure}
    \centering
    \includegraphics[width=0.9\linewidth]{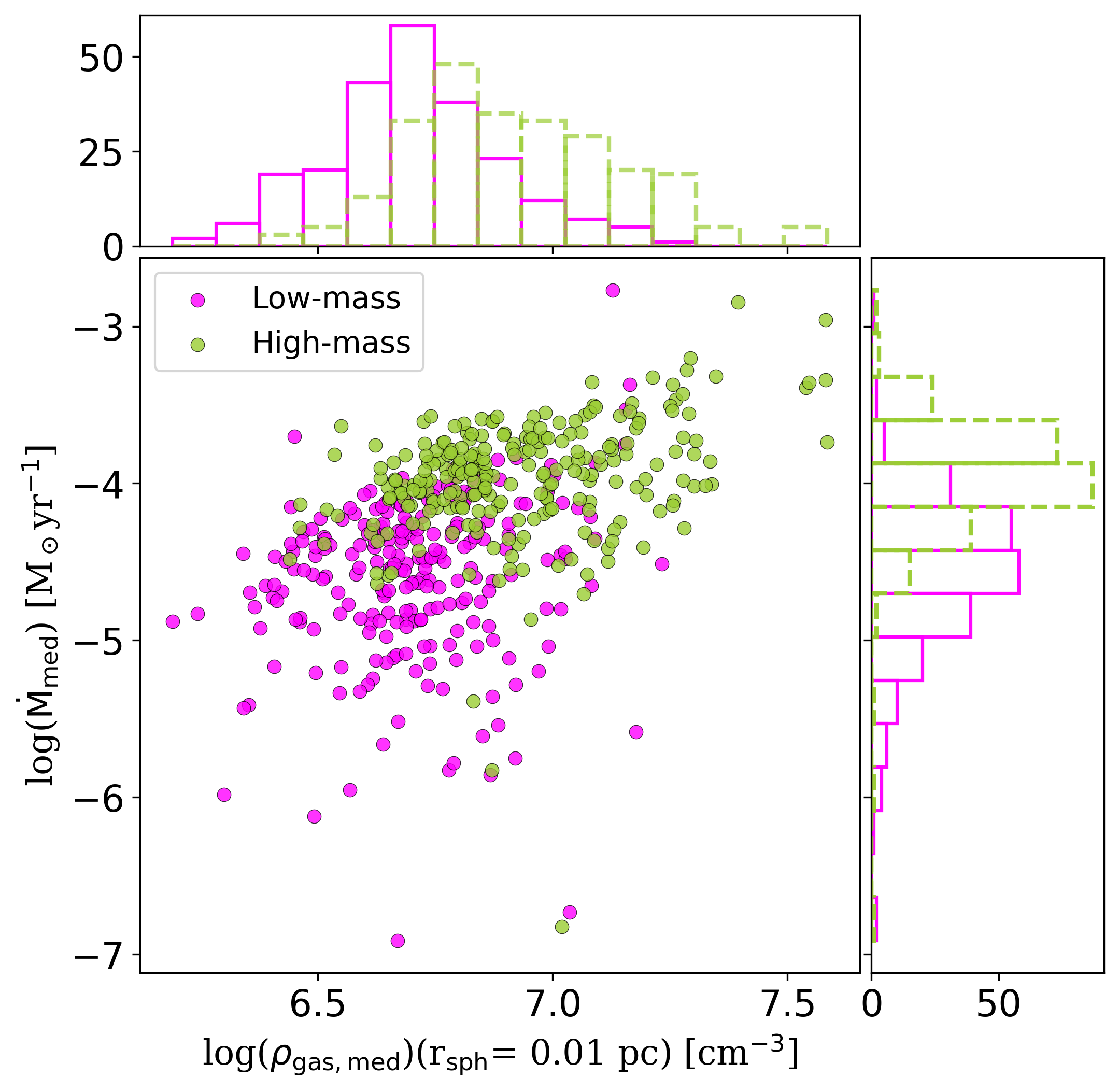}
    \caption{Median mass accretion rate of each sink particle {\it vs.} median gas density around each sink. Gas densities are computed in a sphere of $r_{\mathrm{sph}}=0.01$~pc around each sink. The right and upper panels are the corresponding histograms of the median accretion rate and median gas density around each sink particle. As before, the colors indicate whether the stars are more or less massive than $M_{\rm med, end}$.}
    \label{fig:sim-mdot-gasdens}
\end{figure}

If some particles exhibit larger accretion rates than others, one may expect the density around the stronger accretors to be larger than the density around the lower accretors. In Fig.~\ref{fig:sim-mdot-gasdens} we plot the median mass accretion rate of each sink particle\footnote{We used median (in time) values because the instantaneous accretion rate is highly variable, due to limitations of the numerical simulations \citep[see, e.g., ][]{Ballesteros-Paredes+15, Kuznetsova+18}}, against its median gas density. The color code is as in the previous figures, magenta (green) dots denote the sinks that end up being less (more) massive than the median mass at the end of the simulation, $M_{\rm med,end}$. This figure shows that the sink particles that become more massive have, statistically speaking, larger accretion rates and gas densities. 
The right and upper panels are the histograms of the mean accretion rate and the mean gas density, respectively, around each sink.
As is also clear from this figure, massive stars have, statistically speaking, larger accretion rates and are surrounded by larger gas densities than low-mass stars.

\begin{figure}
    \includegraphics[width=0.9\linewidth]{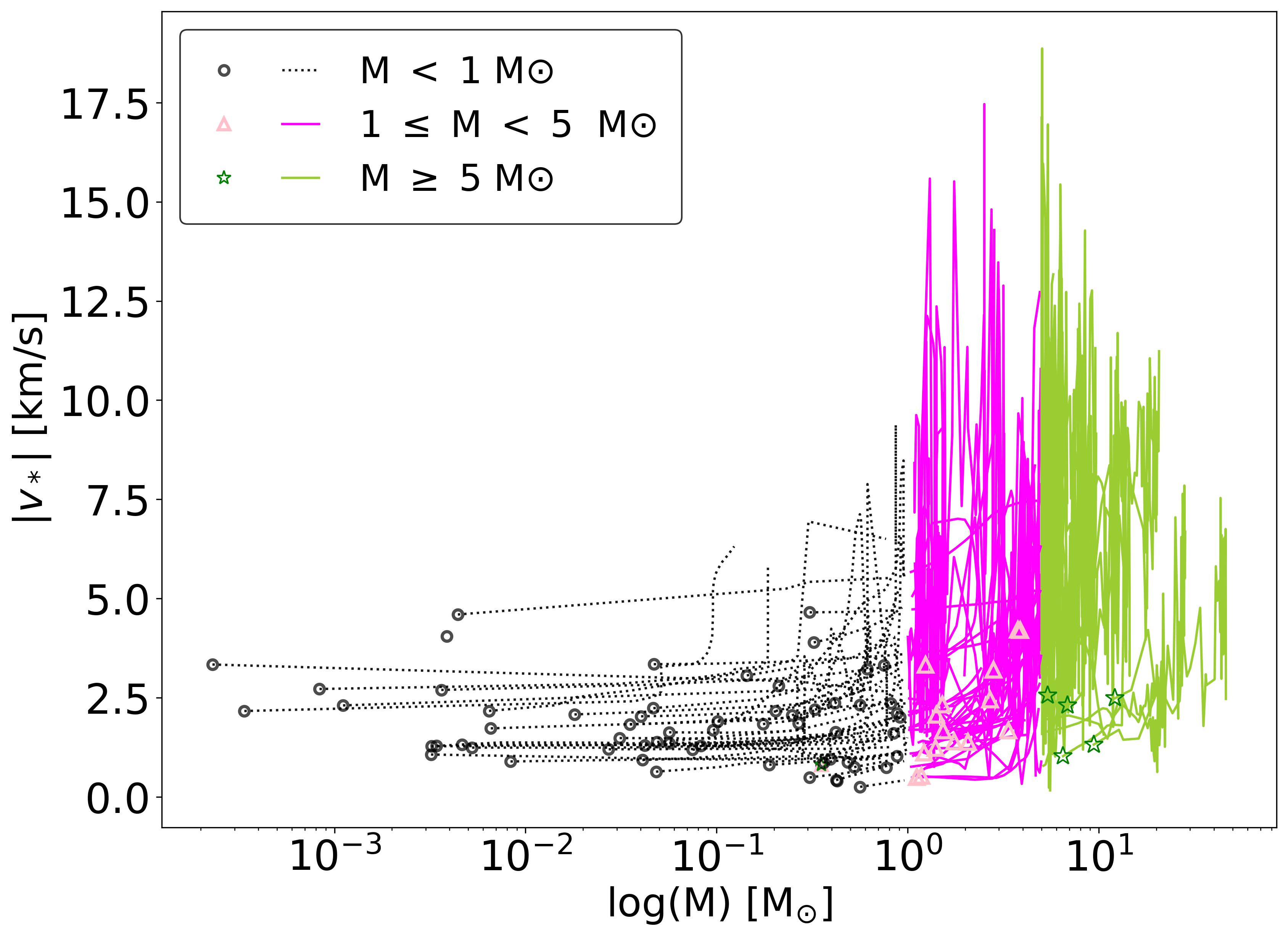}
    \caption{Velocity and mass evolution of the sink particles in group M10S1-G1. The symbols represent the position of each star in this diagram at birth time. The color of the lines denotes the mass range: black for $M < 1$~\Msun, magenta for $1\le M/M_\odot \le5$, and green for $M>5$~\Msun.}
    \label{fig:v-mass:group}
\end{figure}

To investigate how the velocity dispersion of the particles changes as they become more massive, in Fig.~\ref{fig:v-mass:group}, we plot the magnitude of the velocity of each sink as a function of their mass. The lines denote the trajectories of the sinks in the velocity-mass plane. The colors of the symbols denote the sink's masses: black, magenta, and green for the low- ($M\le1~M_\odot$, intermediate- ($1> M/M_\odot\le5$), and high- ($M>5M_\odot$) mass ranges defined at the beginning of this section. The circles, triangles, and stars denote the velocity and mass at the moment of the creation of each sink particle. We notice that the velocities of the lower-mass sinks remain reasonably constant. However, as sinks increase their mass above 0.1~$M_\odot$, their velocities begin to fluctuate substantially, and the larger the masses, the larger the velocity fluctuations. 

\begin{figure}
    \centering
    \includegraphics[width=0.9\linewidth]{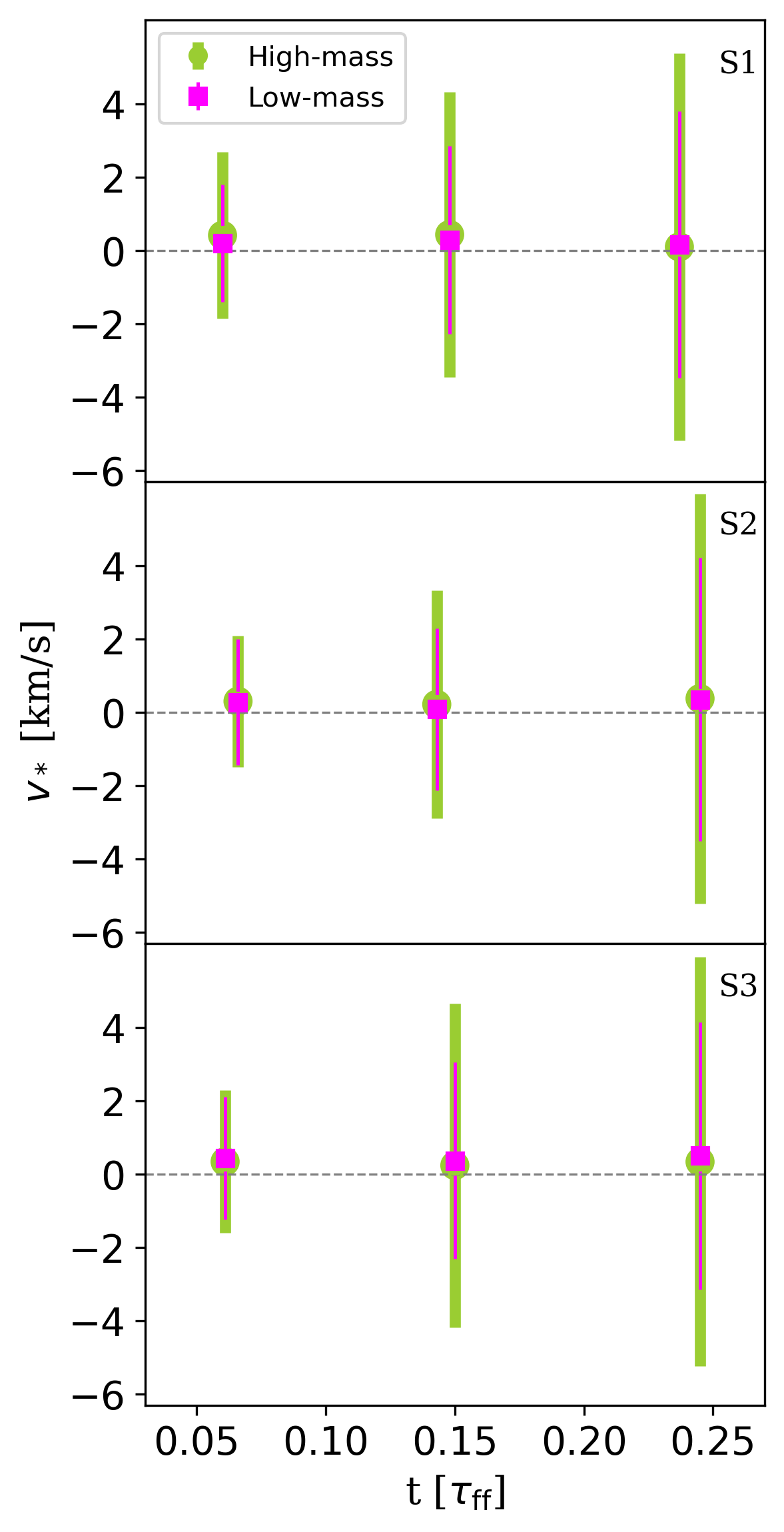}
    \caption{Evolution of the mean velocity (points) and the velocity dispersion ($\pm |\sigma_v|$, error bars) for all sinks in each one of the three simulations, at $t=$~0.06, 0.148 and 0.244~$\tauff$.}
    \label{fig:sinks_vt_bins}
\end{figure}

We also quantify this phenomenon in a different way. In Fig. \ref{fig:sinks_vt_bins} we show the mean velocity (symbols) and velocity dispersion (error bars) of the low- (magenta) and high- (green) mass\footnote{To compare the velocity dispersion between low- and high-mass at different times, we divide low- and high-mass by the {\it instantaneous} median mass, i.e., the median mass of the sample at the corresponding analyzed time.}
stellar particles at times $t=$~0.06, 0.148 and 0.244~$\tauff$. The different panels show the results of our three Mach 10 runs, denoted by S1, S2, and S3. Four features are worth noticing: 

\begin{enumerate}
 
 \item{} At early times, the velocity dispersion of the stars is of the order of $\sigma_v \sim$2~$\kms$, similar to the velocity dispersion of the gas.

 \item{} Although massive stars in M10S1 exhibit a slightly larger velocity dispersion than the low-mass stars at early times, it is unclear whether this trend can be generalized. Indeed, the differences in the velocity dispersion of high- and low-mass stars in runs M10S2 and M10S3 are marginal. As we will see below, our additional runs with different Mach numbers did not allow us to conclude whether the massive stars are born with an initial larger velocity dispersion than low-mass stars. 
 
 \item{} At all subsequent times in all runs, the velocity dispersion of both groups increases.
 
 \item{} The increase of the high-mass stars' velocity dispersion is larger than that of the low-mass ones. 

\end{enumerate}

Point (i) indicates that the sinks formed from turbulent density fluctuations in these M10 runs appear to inherit the velocity dispersion of their parental cloud\footnote{The velocity dispersion is $\sim$2~$\kms$ at early times, which is the velocity dispersion of the gas in our Mach~10 simulations ($T\sim10$~K, isothermal turbulent cloud).} without distinction between the velocity dispersion of massive and low-mass stars. 
Point (ii) shows that we can discard the very first mechanism proposed in the introduction for explaining the larger velocity dispersion of massive stars, namely, that the dynamical heating of massive stars may have a turbulent origin. Indeed, this point shows that there is no distinction between the velocity dispersion of massive and low-mass stars formed from turbulent shocks.
Furthermore, point (iii) strongly suggests that the continuous increase of the velocity dispersion at later times must be somehow related to $n-$body interactions between the sinks as the cloud evolves. Finally, point (iv) shows that such interactions must be stronger for high-mass stars than for low-mass stars. 

\begin{figure}
    \centering \includegraphics[width=0.9\linewidth]{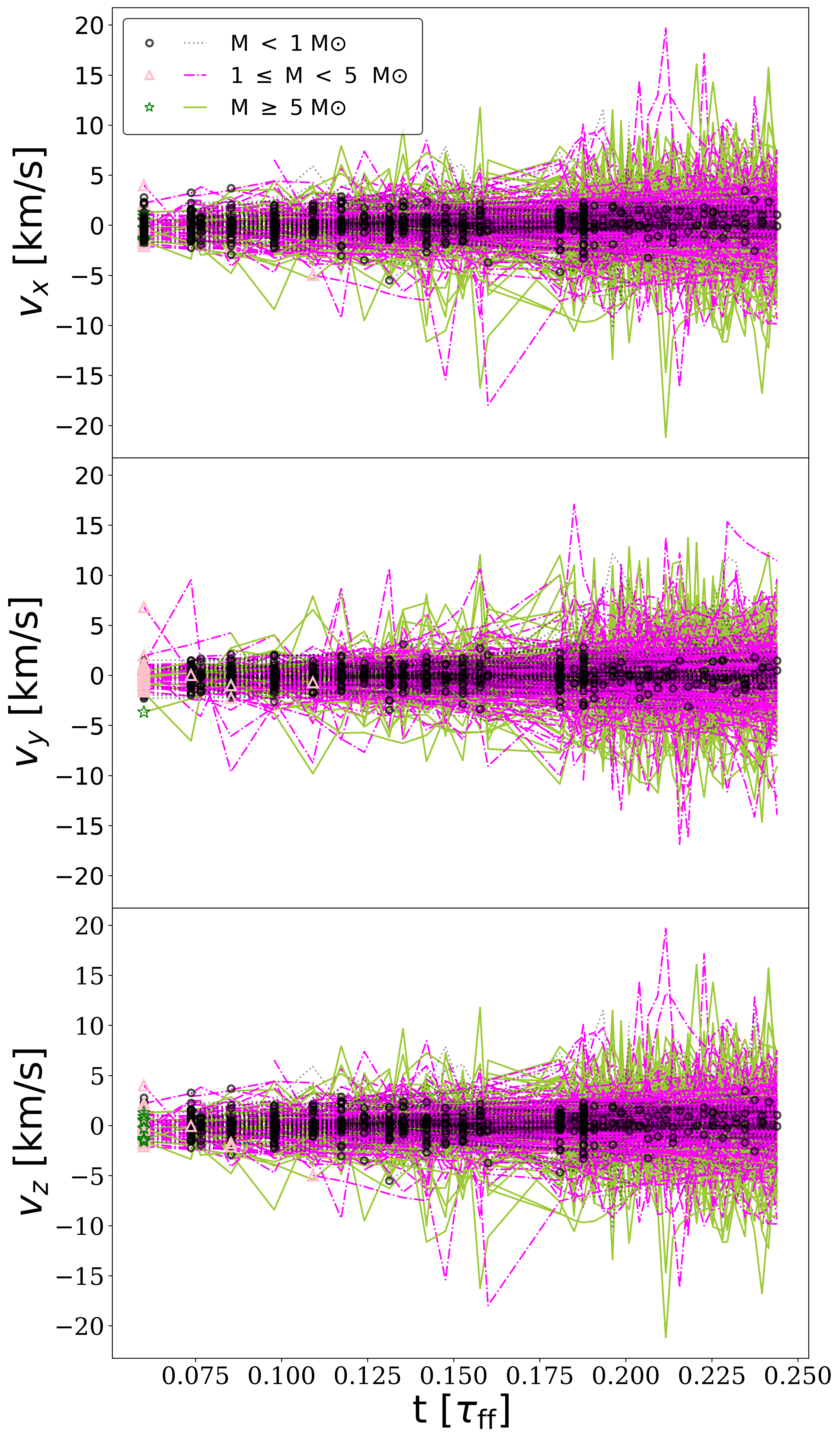}
    \caption{Evolution of the velocity components of each sink in the M10S1 simulation. The symbols are located at the birth-time of each star, whose born mass is in the range of the corresponding colors. The black dotted line shows the evolution in time while the star has $M \le 1$ \Msun, the magenta doted-dashed line shows the changes while the star has 1\Msun $\leq$ M $\le$ 5 \Msun, and the green line the trajectory when M$>$5 \Msun.}
\label{fig:sinks_vt_comps}
\end{figure}

Figure~\ref{fig:sinks_vt_comps} shows in more detail the evolution of the three components of the velocities of all sinks in run M10S1. As in Fig.~\ref{fig:v-mass:group}, the sinks born with masses smaller than 1~\Msun\ are denoted with a black circle; pink triangles denote sinks with masses between $1\le M< 5$ ~\Msun, and green stars denote sinks with masses $M>5$~\Msun. Similarly, dotted black, dotted-dashed magenta, and solid green lines represent the evolution of the sinks in these mass ranges respectively\footnote{As a note of caution for the reader, stars are born at an approximately constant rate in time. The fact that our newborn sinks appear at discrete times is only the result of limiting the dumping of the data. As an extreme case, the lack of newborn stars between $t\sim$0.150 and 0.175$\tauff$ results from some data loss due to a power failure in our supercomputer cluster.}. 
As discussed before, we can notice in Fig.~\ref{fig:sinks_vt_bins} that stars are born with nearly constant velocity dispersion at all times, around $\pm2~\kms$, which is the velocity dispersion of the gas in the simulation. However, the evolution of the individual velocity components in Fig.~\ref{fig:sinks_vt_comps}, indicated by the spread of the lines in the vertical direction, increases with time as the stars become more massive. Interestingly, it is larger for all colors at later times, but the increase is more pronounced for massive stars (green lines) than for intermediate-mass stars (magenta lines), and these are larger than the low-mass stars (black lines), consistent with Figs.~\ref{fig:v-mass:group} and \ref{fig:sinks_vt_bins}.
In addition, we notice that the first sink particles appear soon after gravity is turned on (a feature that can also be seen in Figs.~\ref{fig:sinks_vt_bins} and \ref{fig:sinks_vt_comps}). 
This is a typical result of simulations where turbulence is fully developed when gravity is turned on. In contrast, when gravity is turned on in a homogeneous, turbulent density field, it may take around $t~\gtrsim~3/4~\tauff$ to form the first sinks \citep[e.g.,][]{Ballesteros-Paredes+15, Kim+21}, or even of the order of one free-fall time if the initial shape is filamentary (e.g., Zavala-Molina et al., in preparation).

\begin{figure*}
    \centering
    \includegraphics[width=0.33\linewidth]{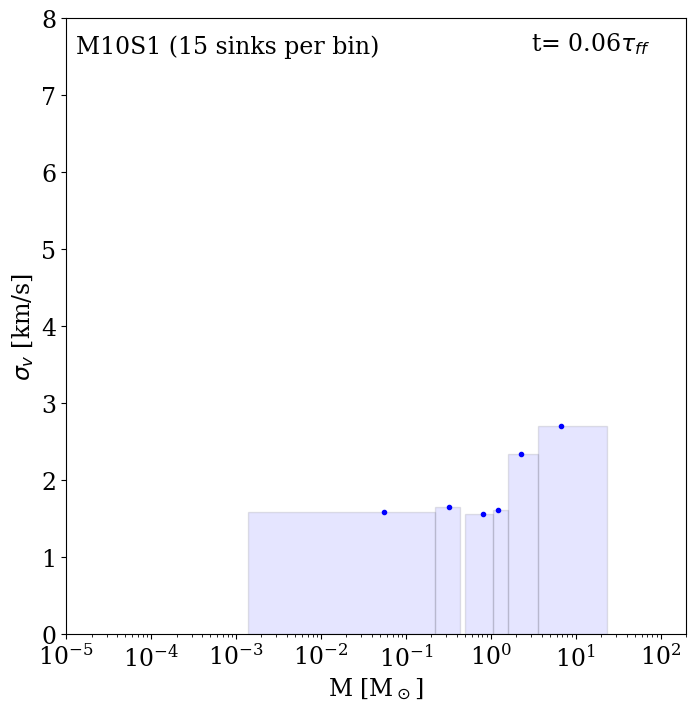}
    \includegraphics[width=0.33\linewidth]{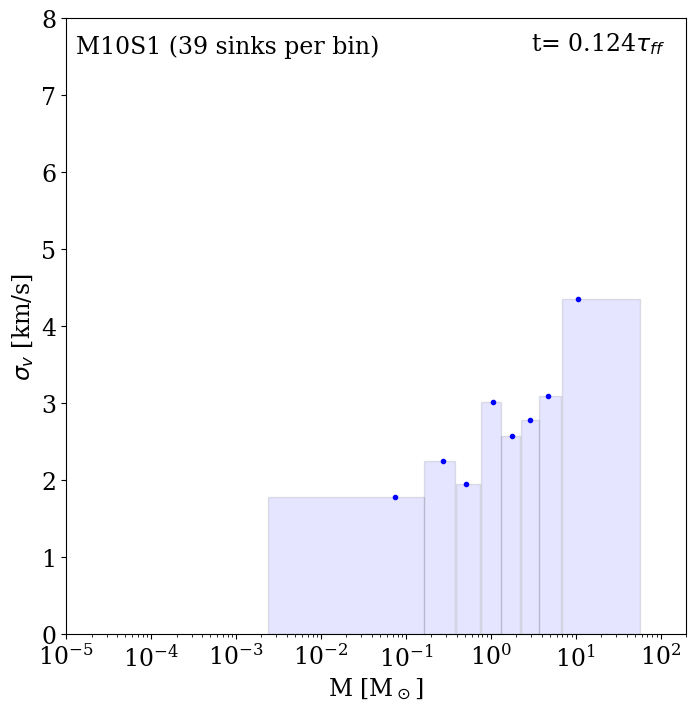}
    \includegraphics[width=0.33\linewidth]{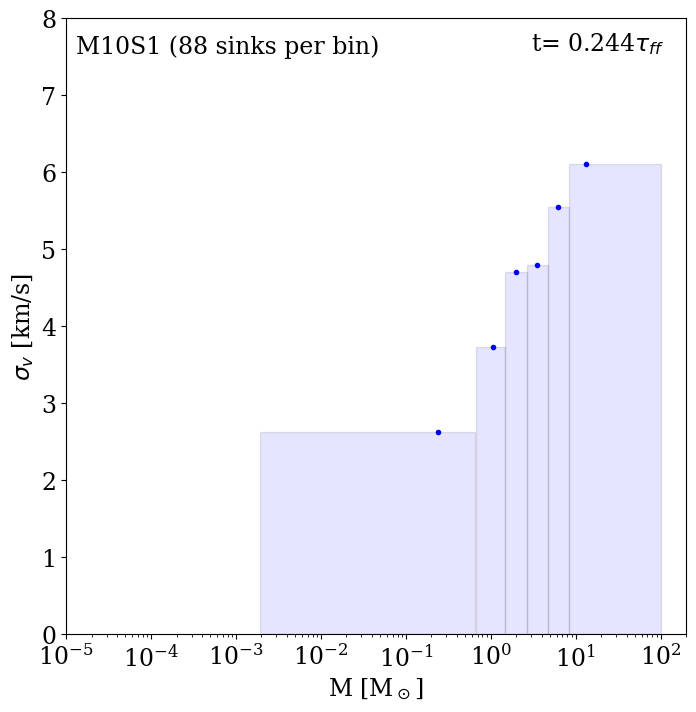}

    \includegraphics[width=0.33\linewidth]{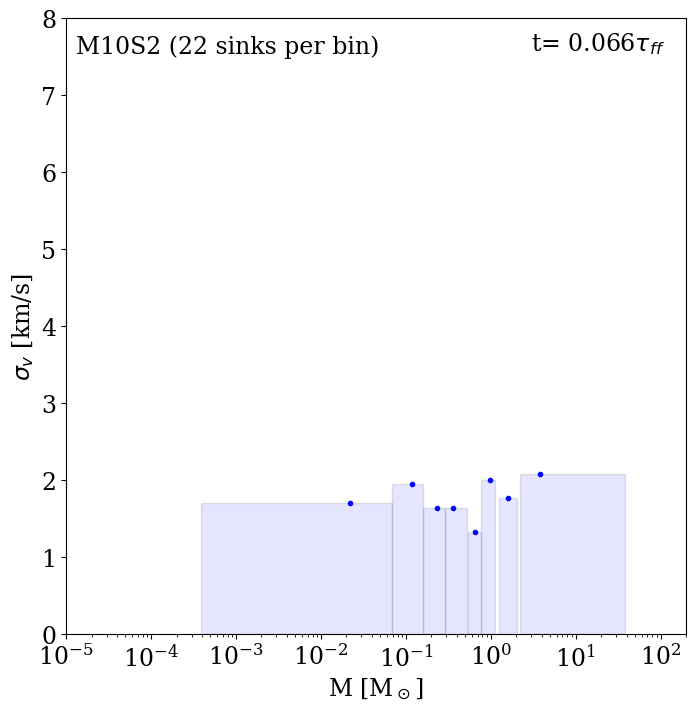}
    \includegraphics[width=0.33\linewidth]{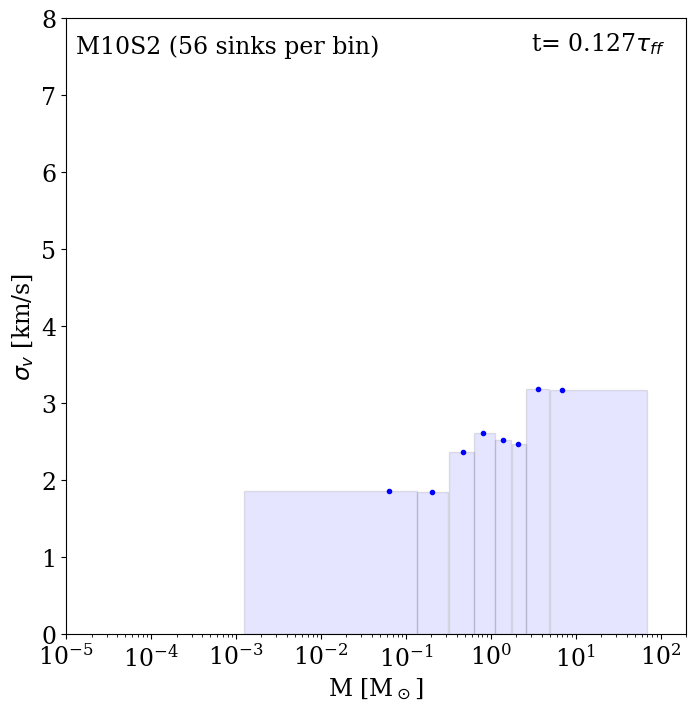}
    \includegraphics[width=0.33\linewidth]{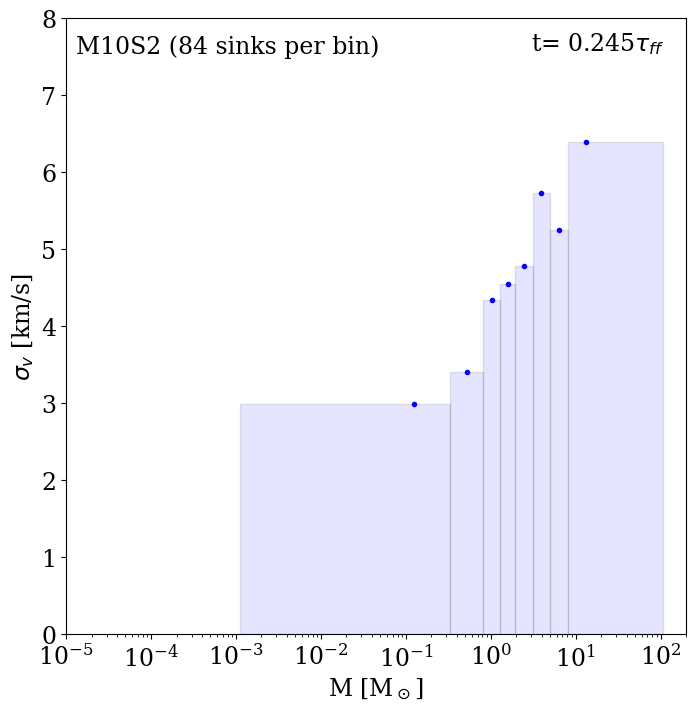}

    \includegraphics[width=0.33\linewidth]{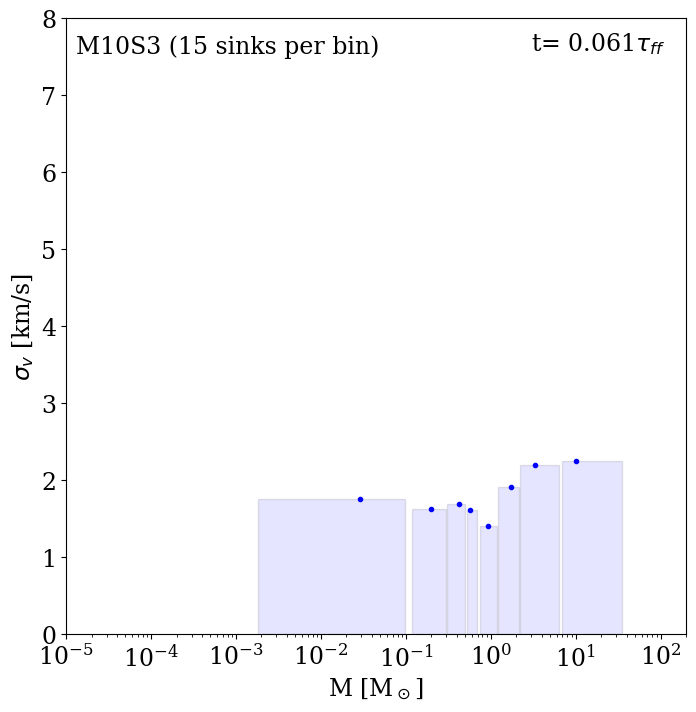}
    \includegraphics[width=0.33\linewidth]{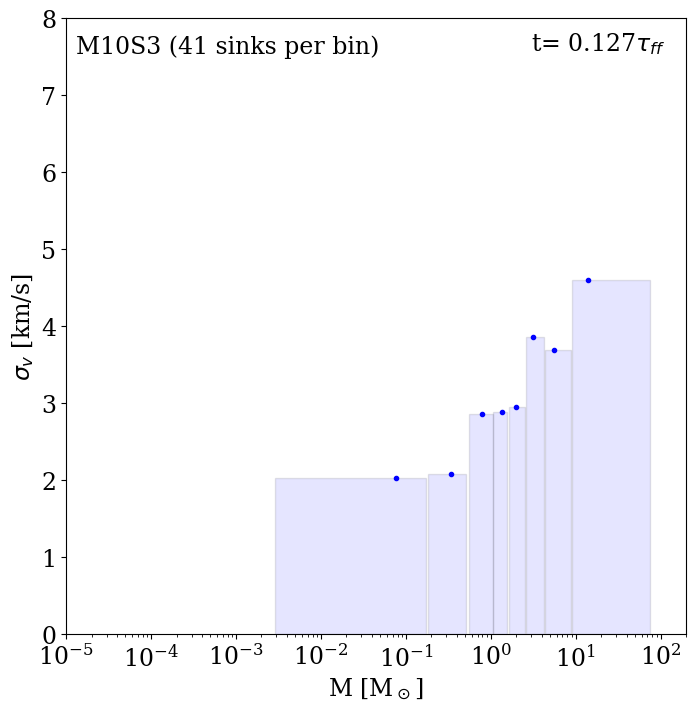}
    \includegraphics[width=0.33\linewidth]{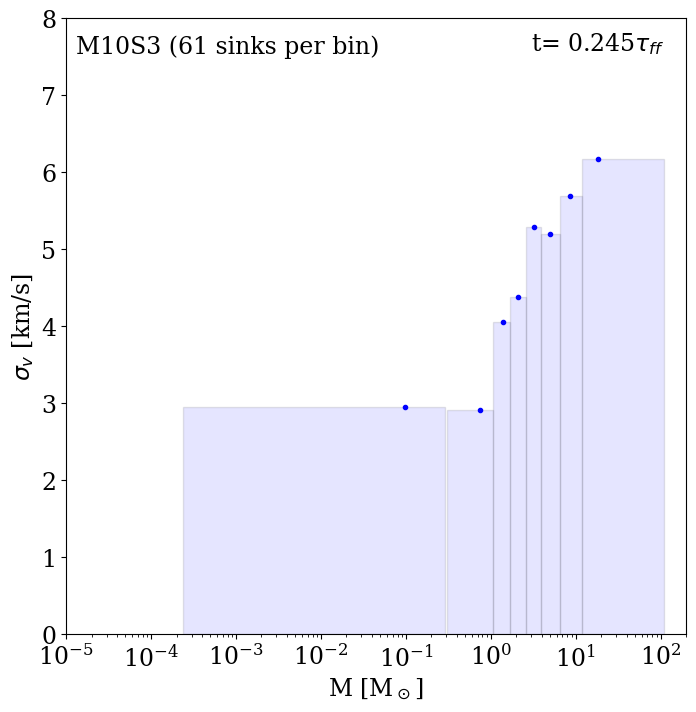}

    \caption{Evolution of the velocity dispersion per mass bin for sinks in turbulent clouds. From the top to the bottom, we show the simulation with different seeds, S1, S2, and S3 respectively. From left to right, the evolution from the first snapshot after gravity is turned on, to the time when they reach $\sim$ 0.245 of the free-fall time. The blue dot in each bin represents the median value for its mass range. As in the large box from \citet{Bonilla-Barroso+22}, each simulation shows a velocity increment with mass.}
    \label{Fig:sigmav-m}
\end{figure*}

Another way to quantify the increase of the velocity dispersion of the sinks as they become more massive is given in Fig.~\ref{Fig:sigmav-m}, where we show the velocity dispersion per mass bin of our sinks in the three Mach 10 simulations, as we did in \citet{Bonilla-Barroso+22} and \citet{Bonilla-Barroso+24} for the Orion and Lagoon Nebula Clusters, as well as for numerical simulations. The reason for making this plot with a histogram representation is that we want to explicitly show the ranges of the mass bins used to compute the velocity dispersion. {\color{black}It should be noticed, as denoted in each frame, that all bins in a single plot have the same number of sinks and thus have the same statistical significance.} From top to bottom, the figure shows run M10S1, M10S2, and M10S3, respectively. From left to right, we show the beginning, middle, and final timesteps of the studied period ($t=$~0.06, 0.148 and 0.244~$\tauff$, respectively). It is clear from this plot that the more massive the stars become, the larger their increase in their velocity dispersion, consistent with Figs.~\ref{fig:v-mass:group}, \ref{fig:sinks_vt_bins}, and \ref{fig:sinks_vt_comps}. 

\begin{figure}
    \centering
    \includegraphics[width=0.74\linewidth]{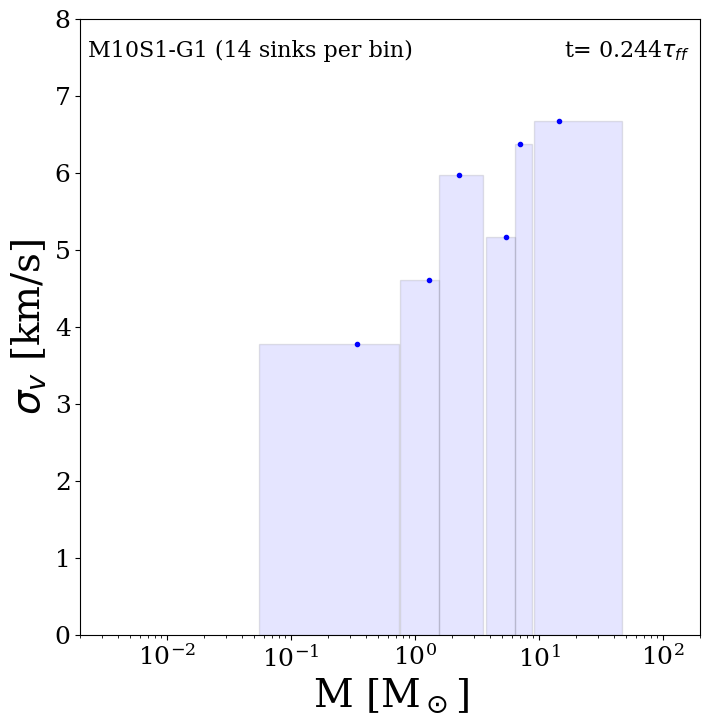}  
    \caption{Velocity dispersion per mass bin at $t\sim 0.25 \tauff$ for the selected stellar group in M10S1.}
    \label{fig:non-histo-group}
\end{figure}

To understand the physical cause of the systematic increase in the velocity dispersion of high-mass stars with time, we proceed to analyze in detail the group M10S1-G1. The results for other groups are similar to what we present here. 
First of all, similar to Fig.~\ref{Fig:sigmav-m}, in Fig.~\ref{fig:non-histo-group} we plot the velocity dispersion {\it vs.} mass for the stars, but only for the group M10S1-G1 at the final time. As expected, the sinks in the individual groups exhibit the same behaviour as before for the whole simulation, i.e., the massive stars develop velocity dispersions larger than the low-mass stars.\\

\begin{figure*}
    \includegraphics[width=0.9\linewidth]{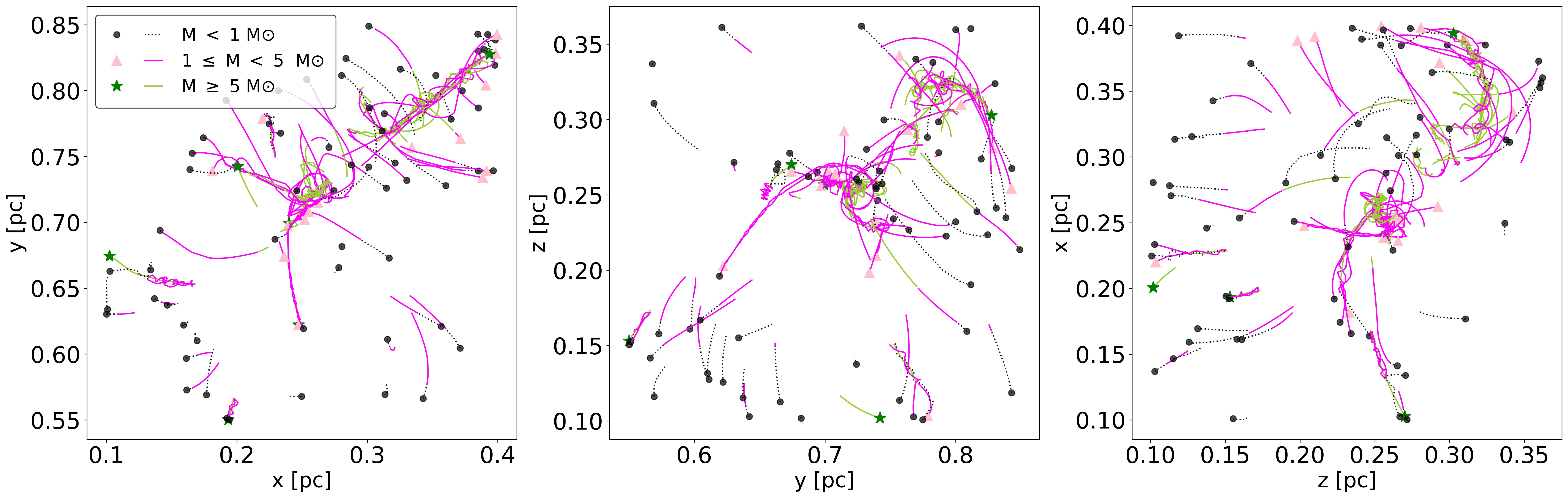}
    \caption{Stellar particle trajectories for a group in M10S1 simulation (selected region in Fig.~\ref{fig:coldens_s1}) during the studied period $\sim 0-0.25 \tauff$. The symbols show the position when a sink particle first appears and the lines show its path through its evolution. The terminology is the same as in Fig. \ref{fig:coldens_s1} (see also the inner box in the left panel). }
    \label{Fig:group-pp}
\end{figure*}

In Fig.~\ref{Fig:group-pp}, we now show projections in $x-y$, $y-z$, and $z-x$ of the trajectories of the stars for the studied group. The symbols and color codes are the same as in the previous figures showing the three mass ranges defined at the beginning of this section. As can be noticed, most stars are born with low masses (black dots and dotted black trajectories).
Additionally, it can be observed that, as time passes, the sinks that are in regions with larger density of stars,  
(a)~have a larger increase in their masses (as indicated by the color code), suggesting higher accretion rates, and 
(b)~interact with their closer neighbours, as denoted by the tightening and entanglement of their orbits. 
(c)~This tightening and entanglement naturally explain the increasing velocity dispersion, due to closer, more bounded orbits. 
In contrast, the sinks that are formed in less crowded regions 
(i) remain typically less massive, suggesting lower accretion mass rates, and their orbits remain less entangled, suggesting that 
(ii) their velocity dispersion does not increase substantially.

\begin{figure}
    \centering
    \includegraphics[width=0.9\columnwidth]{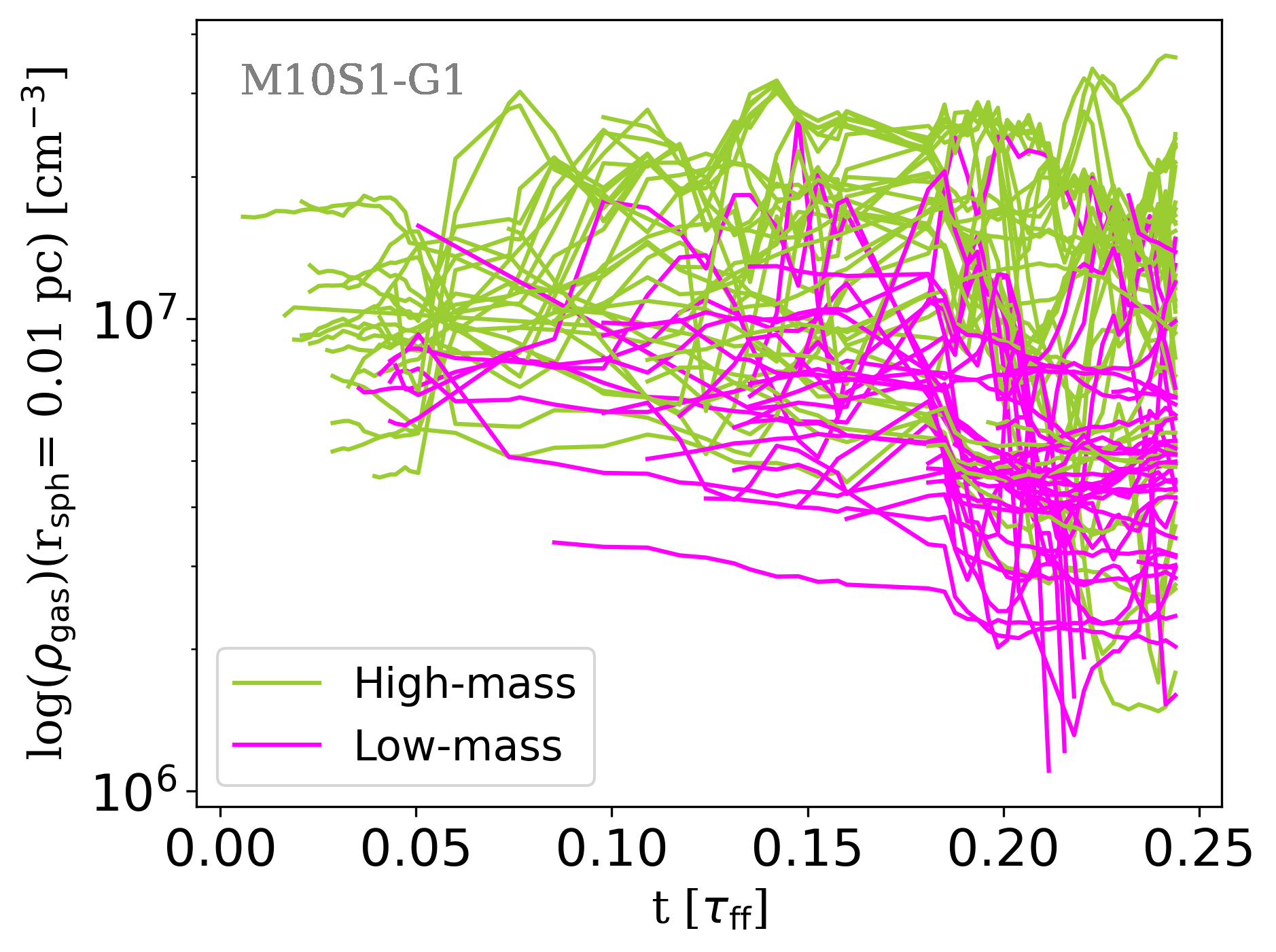}  
    \caption{Temporal evolution of gas density for each stellar particle. These densities are computed within a spherical volume of radius $r_{\mathrm{sph}}=$~0.01~pc around each particle. From this figure, it is clear that stars that end up being more massive (green lines) tend to live in denser regions than the stars that end up being less massive (magenta lines).}
    \label{fig:density-time}
\end{figure}

The previous analysis suggests that the stars increase their velocity dispersion because their orbits become tighter as they increase their mass. To quantify further, in Figure \ref{fig:density-time} we show the evolutionary track of the mean gas density, measured inside a spherical volume of radius $r=$~0.01~pc around each sink. As it can be noticed (see also Fig.~\ref{fig:sim-mdot-gasdens}), the gas density is larger for those stars ending up with larger masses.

Additionally, in  Fig.~\ref{fig:histogram_neighbours2}, we show the histogram of the number of neighbours $N_{\rm ngh}$ within a radius of $r=$~0.01~pc, for all sinks at all times. As before, the green histogram denotes the stars ending with masses above the median mass $M_{\rm med, end}$, whereas the magenta histogram denotes stars ending with masses below it. It is clear from this histogram, that the more massive stars have, statistically speaking, a larger number of companions. 

\begin{figure}
    \centering
    \includegraphics[width=0.92\linewidth]{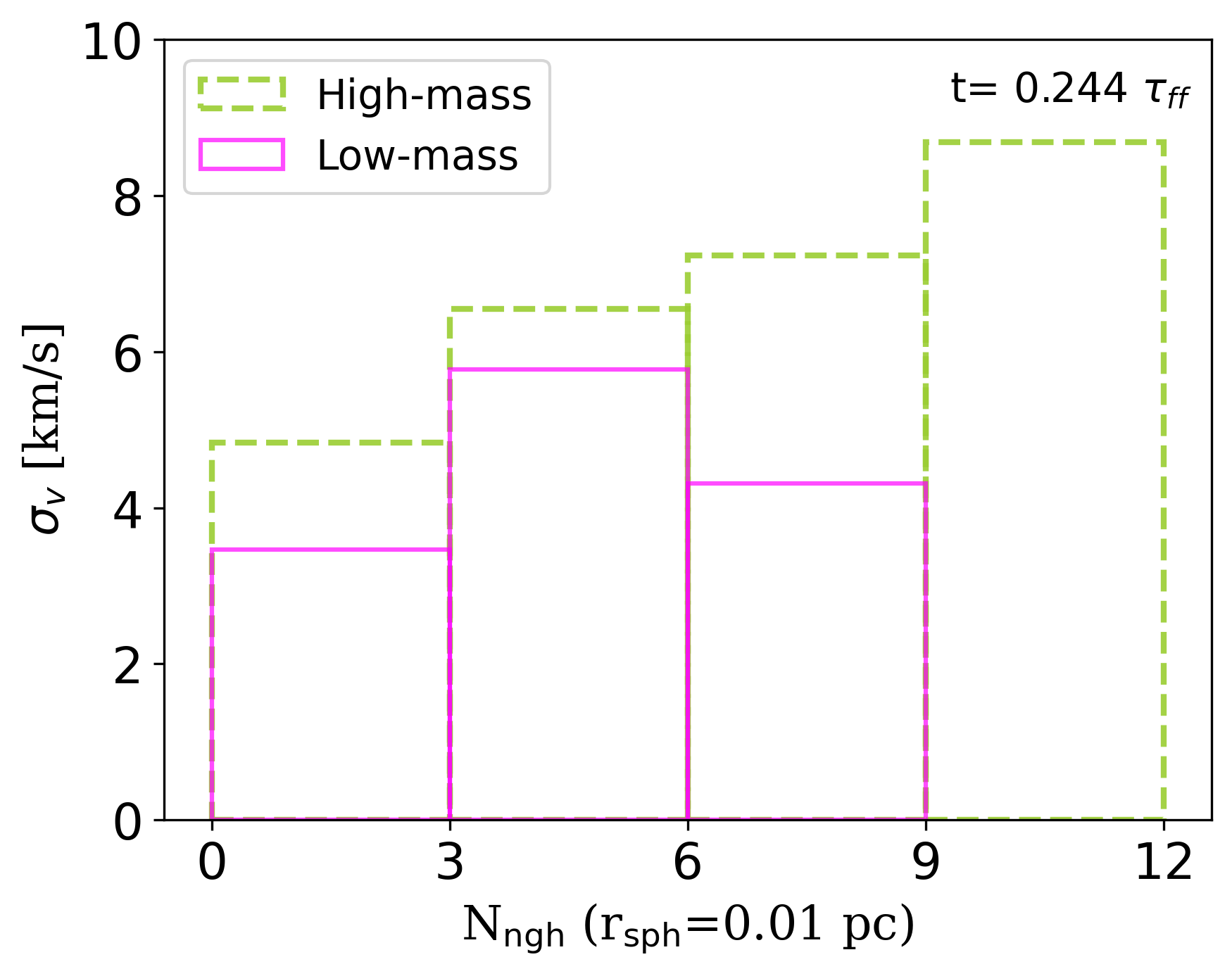}
    \caption{Velocity dispersion of stars in model M10S1 with N$_{\rm ngh}$ neighbours within a spherical radius $r_{\rm sph}=$~0.01~pc, at $t=0.244~\tauff$.  It is clear from this figure that the massive stars exhibit larger velocity dispersion, but also that, statistically speaking, tend to have more neighbours.}
    \label{fig:sigmaV-Nneigh}
\end{figure}

Finally, in Fig.~\ref{fig:sigmaV-Nneigh}, we show the velocity dispersion of the sinks as a function of the number of neighbours, showing that the larger the number of neighbours, the larger the velocity dispersion of the stars. As can be seen, the massive stars tend to have a larger number of companions, and their velocity dispersion increases as the number of companions increases.  \\

In summary, our results on stellar kinematics in turbulence-supported clouds show that: 

\begin{enumerate}

 \item{} Stars are born either in a grouped, or isolated fashion (Figs.~\ref{fig:coldens_s1} and \ref{Fig:group-pp}). 

 \item{} Statistically speaking, stars that grow to higher masses exhibit larger mass accretion rates than less massive stars (Figs.~\ref{fig:accrete} and \ref{fig:sim-mdot-gasdens}). 
 
 \item{} As stars become more massive, they increase their velocity dispersion (Figs.~\ref{fig:v-mass:group}, \ref{fig:sinks_vt_bins}, \ref{fig:sinks_vt_comps}, \ref{Fig:sigmav-m} and \ref{fig:non-histo-group}). 
 
 \item{} Grouped stars tend to become more massive and to develop more entangled orbits (Fig.~\ref{Fig:group-pp}) than isolated stars. 
 
 \item{} Stars that evolve to higher masses are located in volumes with larger gas and stellar densities than those that remain less massive
 (Figs.~\ref{fig:sim-mdot-gasdens}, \ref{fig:density-time} and \ref{fig:histogram_neighbours2}). 

 \item{} Massive stars tend to have a larger number of neighbours and larger velocity dispersion than low-mass stars (Fig.~\ref{fig:sigmaV-Nneigh}).

\end{enumerate}

In the next section, we discuss in detail these results and their implications in our understanding of the physical causes of the dynamical heating of massive stars.

\section{Discussion}\label{sec:discussion}

It has been shown previously that stars formed in collapsing clouds develop a constant velocity dispersion, regardless of the mass range of the stars \citep{Bonilla-Barroso+22, Bonilla-Barroso+24}. This behaviour is a natural consequence of the violent relaxation that occurs during the process of collapse \citep{Lynden-Bell67}, in which the change in the gravitational potential dominates the forces of the particles in the system. Since the accelerations over each particle are provided by the total gravitational potential rather than individual interactions between the stars, and since the potential varies with time, the stars develop a velocity dispersion that does not depend on their mass.

In contrast, as shown in the previous sections, stars formed in globally supported supersonic turbulent clouds develop a velocity dispersion that depends on their mass. Although we found this result in an earlier contribution \citep{Bonilla-Barroso+22}, in that work, we did not verify the cause of such dynamical heating of the massive stars. When we started this contribution, we envisioned three possible explanations: a turbulent origin, a collisional relaxation mechanism, and accretional tightening of the orbits of the stars.  
In what follows, we discuss each of these scenarios and whether our findings favor one or another possibility.

\subsection{Turbulent case}

We envisioned this as a possible mechanism in one of our previous contributions \citep{Bonilla-Barroso+22}. The idea was that massive cores are, statistically speaking, denser. Since in a turbulent environment, stronger shocks form denser regions than less intense shocks \citep[e.g.,][]{Vazquez-Semadeni94, Padoan_Nordlund02}, one could expect that the core-to-core velocity dispersion could be larger for massive clumps than for low-mass cores because, in a way, cores must inherit the velocity dispersion of the shock that produces them. Furthermore, since high-mass stars are supposed to be formed in massive cores and low-mass stars in less massive cores, one could expect massive stars to have a larger velocity dispersion than low-mass stars.

From Fig.~\ref{fig:sinks_vt_bins}, it became clear that this is not the case, there is no substantial difference between the velocity dispersion of the newborn high- and low-mass sinks right after self-gravity is turned on, implying that both types of stars are born from cores with similar velocity dispersion.

\subsection{Collisional relaxation}

Another possibility was that collisional relaxation is responsible for massive stars having larger velocity dispersion than low-mass stars. The idea of this mechanism is simple. The tendency towards energy equipartition in $n-$body systems induces high-mass stars to transfer energy to low-mass stars. However, since gravitational systems have a negative heat capacity \citep[e.g.,][]{Lynden-Bell_Wood68, Antonov85}, this results in the opposite effect, with high-mass stars ending up with {\color{blue} a} larger velocity dispersion {\color{blue}than low-mass stars}. This occurs because, as high-mass stars give net energy to low-mass stars during $n-$body interactions to achieve equipartition, they tend to fall into the local potential well of the group. As this occurs, they increase their velocity dispersion. In contrast, low-mass stars gain net energy, which allows them to move to less-bound, more elongated orbits of their local group, lowering their velocity dispersion. Consequently,  high-mass stars end up with a larger velocity dispersion than low-mass stars. An extreme case of this situation is the \citet{Spitzer69} {\color{blue}{instability}}, where massive stars dominate the gravitational potential at the cluster's center. Thus, they evolve at their own dynamical time, ejecting even massive stars from the cluster's center. 

The analyses of our turbulent simulations performed in the current contribution show that this is not the case either. For this mechanism to have played a role, we should have found a substantial fraction of low-mass stars expelled from the densest regions of the groups into less-bound orbits after several-body encounters with more massive stars. Although this phenomenon happens, this is not the primary mechanism occurring in our simulations.

\subsection{Accretional tightening}

\begin{figure}
    \centering\includegraphics[width=0.95\linewidth]{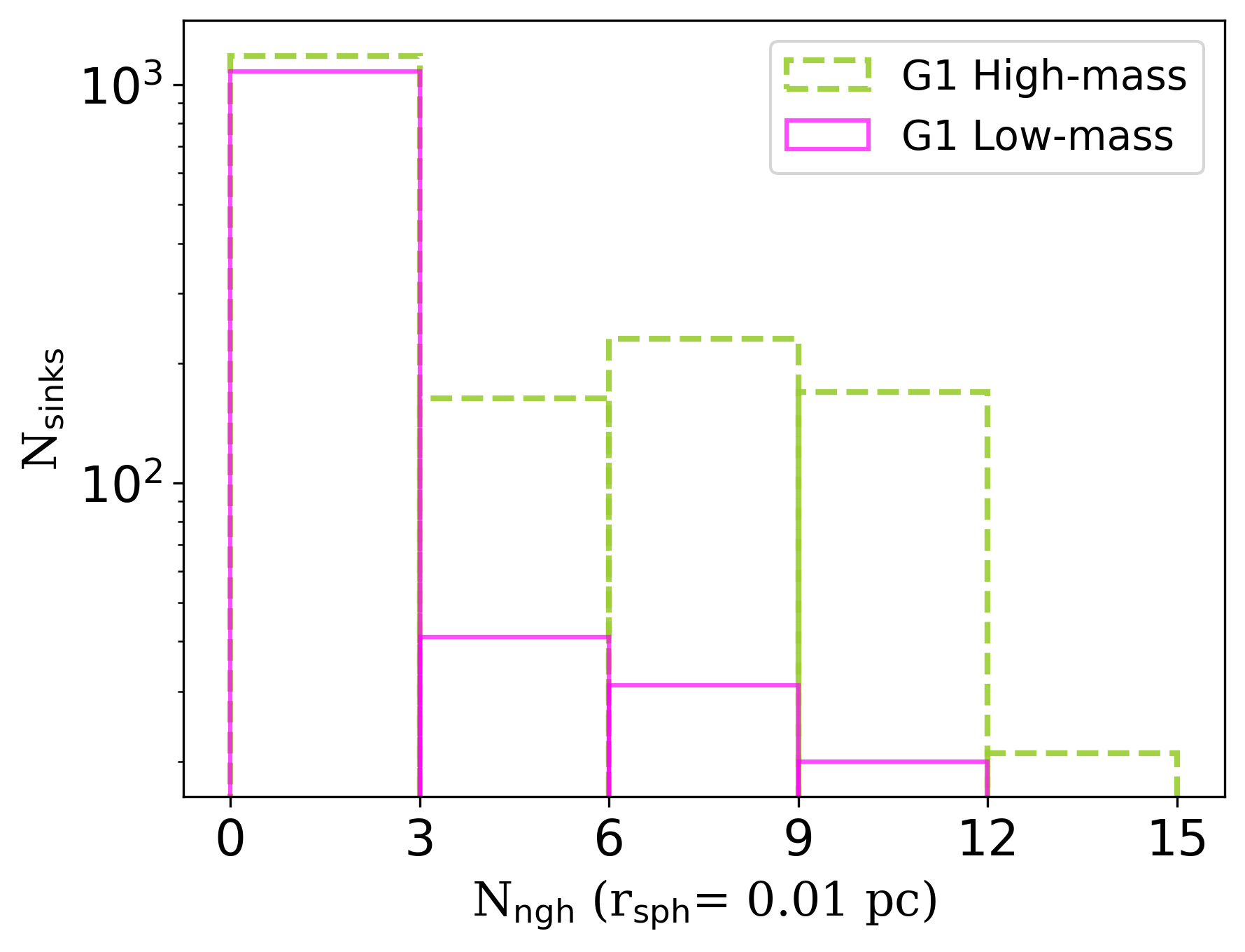}
    \caption{Histogram of the number of neighbours N$_{\rm{ngh}}$ inside the sphere of $r=$~0.01~pc for particles at all times in the M10S1-G1 group. As can be seen, more massive sinks tend to have a larger number of companions.}
    \label{fig:histogram_neighbours2}
\end{figure}

Our simulations show an interesting behaviour. Almost all sinks start as low-mass stars (Figs.~\ref{fig:v-mass:group}, and \ref{Fig:group-pp}). As they increase their masses, their orbits become more entangled, especially those in more crowded places (Fig.~\ref{Fig:group-pp}). While this happens, they also increase their velocity dispersion, with high-mass stars having a larger increase (Figs.~\ref{fig:v-mass:group}, \ref{fig:sinks_vt_bins}, \ref{fig:sinks_vt_comps}, \ref{Fig:sigmav-m} and \ref{fig:non-histo-group}). 
Interestingly, the high-mass stars are located preferentially in denser places (Fig.~\ref{fig:density-time}), which explains their larger accretion rates (Figs.~\ref{fig:accrete}, \ref{fig:sim-mdot-gasdens}). By the same token, they are located in more crowded places (Figs.~\ref{Fig:group-pp} and \ref{fig:histogram_neighbours2}). In contrast, many low-mass stars tend to be born in less-dense (Fig.~\ref{fig:density-time}), and less-crowded places (Fig.~\ref{fig:histogram_neighbours2}). The orbits of these stars are not as entangled (Fig.~\ref{Fig:group-pp}), and their velocity dispersion does not increase that much (Figs.~\ref{fig:v-mass:group}, \ref{fig:sinks_vt_bins}, \ref{fig:sinks_vt_comps}, \ref{Fig:sigmav-m} and \ref{fig:non-histo-group}) compared to high-mass stars.

The phenomenology described above occurs not only in group M10S1-G1, but in all our turbulent simulations, indicating that massive stars increase their velocity dispersion more prominently than low-mass stars because they become more gravitationally bound, with closer orbits, as they accrete more mass from their mass reservoir. We call this process {\it accretional tightening.}

\subsection{Implications for models of molecular cloud dynamics}

Turbulence in star-forming molecular clouds has been extensively studied by different authors \citep[see reviews by, e.g.,][and references therein]{Vazquez-Semadeni+00, Ballesteros-Paredes+07,  Hennebelle_Falgarone12, Dobbs+14}. Less well-studied have been the implications of molecular cloud dynamics on the kinematical signatures of stellar systems within their parental cloud as they are formed. Recently, significant effort has been made to put together $n-$body and hydrodynamical codes, in addition to stellar evolution, to follow the evolution of an $n-$body cluster as it forms from its parent cloud \citep[][]{Pelupessy+13, Wall+19, Wall+20}. However, due to computational difficulties in following the evolution of gas and stars simultaneously, most of the studies of stellar dynamics of star-forming regions deal with stellar systems but with no gas in the system \citep[e.g.,][]{Allison+09, Parker_Dale13, Parker+14, Spera+16}.

In previous contributions we followed the early dynamical evolution of the stars within their parental cloud  \citep{Bonilla-Barroso+22, Bonilla-Barroso+24}. We noticed that the stellar kinematics of young clusters can give clues on the actual dynamic state of their parent molecular cloud. Indeed, a constant velocity dispersion of stars as a function of their mass can be understood as the result of the cluster being formed by the collapse of their parent clump. We found this to be the case of either numerical simulations of collapsing clouds and of the Orion and the Lagoon Nebula Clusters.

In contrast, the current study shows that turbulence-supported clouds develop a different velocity dispersion-mass relation, with massive stars having a larger velocity dispersion than low-mass stars.  
Strictly speaking, the dynamical heating of massive stars in the current work is not caused by collisional relaxation or its extreme case, the \citet{Spitzer69} instability, as some authors have suggested \citep{Spera+16, Wright_Parker19}. 
Instead, we observe that it results from the enhanced boundness of stellar groups through accretional tightening of mass from their parental clump. Being larger the accretion of gas mass by massive stars, these are also the stars that undergo larger accretional tightening, and thus, larger dynamical heating.

It should be noticed that, although in turbulent models, the cloud is globally supported by turbulence, local groups are formed within cores undergoing local collapse. Thus, one can ask whether these stars should or should not undergo violent relaxation. The answer is that, for violent relaxation to occur, the local free-fall time of the gas must be smaller than the collisional relaxation time, $\tauff < \taurelax$. In the case of the Orion Nebula Cluster, for instance, we can consider the mean density of the core behind the nebula to be of the order of $n\sim 10^4$~cm\alamenos3, which implies a free-fall time of the order of $\tauff\sim3\times10^5$~yr. As for the stellar cluster, the number of stars is of the order of 2,000, with a velocity dispersion of $\sigma_{\rm v,*}\sim2~\kms$, in a core of size $R\sim1$~pc. Thus, the dynamical crossing time is of the order of $5\times10^5$~yr, and its collisional relaxation time is of the order of 
\begin{equation}
  \taurelax \sim \frac{0.1 N}{\ln{N}}\taudynstars \sim 1.3\times10^7~\mathrm{yr}
\end{equation}
much larger than the free-fall time of the core, explaining why we observe violent relaxation in the Orion nebula cluster \citep{Bonilla-Barroso+22}. In our region M10S1-G1, in contrast, the dynamical time of the stars is, initially, of the order of 
\begin{equation}
 \taudynstars \sim \frac{0.3~\mathrm{pc}}{2~\kms} \sim 1.5\times10^5~\mathrm{yr}
\end{equation}
(where we have used the size of group~1, see Fig.~\ref{fig:coldens_s1}; and initial velocity dispersion of the sinks to be of the order of 2~$\kms$, see Fig.~\ref{fig:sinks_vt_bins}), and becomes smaller as the groups increase their velocity dispersion. With individual stellar groups of $N\sim10$ members, this implies a relaxation time of the order of
\begin{equation}
 \taurelaxstars \sim \frac{0.1~N}{\ln{N}}\taudynstars \sim 6.4\times10^4~\mathrm{yr.}
\end{equation}
In contrast, the free-fall timescale is roughly three times larger, assuming a mean gas density in those groups to be of the order of $2-3\times 10^4$~cm\alamenos3. This longer free-fall time explains why the interactions between the stars dominate the velocity dispersion rather than the collapse itself and why accretional tightening takes place. 

Our results have important implications for the adequate modelling of star-forming clouds. On the one hand, we found that accretion can be important in turbulent boxes. This is at odds with the idea that stars accrete only in a Bondy-Hoyle-Lyttleton fashion, as suggested previously \citep{Krumholz+05}. We do not claim that our accretion scheme is a better or worse representation of the accretion into the stellar groups and finally into the stars. However, we claim that more careful analysis of the accretion of gas mass into the stars has to be performed to follow the cluster formation process in different environments and to understand the gravitational cascade of mass from the parental cloud down to the stars. 

It is interesting to notice that \citet{Offner+09} found that the velocity dispersion of their stellar particles is substantially smaller ($\sim1/5$) than the velocity dispersion of the parental cloud, regardless of whether their clouds are collapsing or turbulence-supported. This result may be skewed, in fact, by their numerical treatment of accretion. Indeed, the accretion scheme of those authors is inherited from the code of \citet{Krumholz+04}, which assumes that accretion into the sinks occurs at a Bondi-Hoyle rate. This implies that the sinks accrete by considering only the gravity of the sink, with an effective cross-section given by the relative velocity between the star and the gas cell. This neglects the gas cell's self-gravity and the possibility of spherical accretion into the sink. As a consequence, a lower accretion rate will be less relevant in increasing the velocity dispersion of the sink groups, explaining the low-velocity dispersion of \citet{Offner+09}. 

Another implication of our work is that if stars change their masses significantly during their early evolution, this process has to be taken into account to understand the kinematical features of stellar clusters and to what extent young stellar groups and clusters remain bound. Schemes of star formation that form stars with a prescription for their masses but do not allow further accretion into individual stars \citep[e.g.,][]{Colin+13, Wall+19} might not be able to follow this effect.\\

Finally, the $n-$body algorithm we use here \citep[from][]{Federrath+10b} may not accurately follow the dynamics of systems bounded below our resolution ($\Delta x \simeq 391$~AUs). Thus the dynamical heating we measure here may be underestimated. However, we also do not include stellar feedback, which may limit accretion, and thus, accretional tightening. Those effects will be considered in a further contribution

\section{Conclusions}\label{sec:conclusions}

In this work, we investigate the causes of the dynamical heating of massive stars in young stellar groups formed in numerical simulations of supersonic turbulent clouds. 
Our study shows that massive stars develop a higher velocity dispersion than low-mass stars, which is consistent with our previous work. This contrasts with stars formed in collapsing clouds, where the velocity dispersion does not depend on the mass of the stars. We discussed three possibilities to explain this behaviour: turbulence, collisional relaxation, and accretion-driven orbital tightening.
Our results support the third scenario, showing that:

\begin{enumerate}

    \item High-mass stars are born in the densest regions of turbulent molecular clouds, while low-mass stars do it mainly in less dense regions. \\

    \item All stars experience accretion, and if they are orbiting each other, their orbits become more bound due to the increasing mass of the particles. Consequently, their velocity dispersion increases.\\
    
    \item Since high-mass stars are formed in the densest parts of the clouds, they tend to come together in groups and to accrete more mass simultaneously. This results in massive stars tightening their orbits and increasing their velocity dispersion more vigorously than low-mass stars, which are more isolated, located at lower gas densities and less-crowded regions.

\end{enumerate}

\section*{Acknowledgements}

The authors acknowledge the anonymous referee for the comments and suggestions that helped to improve the clarity of this manuscript. 
We thank Lee W. Hartmann for helpful discussions regarding the content of this work.
This work has extensively used the SAO-NASA Astrophysical Data System (ADS). 
The numerical simulations were performed in the clusters ``Miztli'', at DGTIC-UNAM, through proposal LANCAD-UNAM-DGTIC-070; and ``Mouruka'', at IRyA-UNAM, provided by CONAHCYT through grant number {\tt INFR-2015-01-252629.}
V.C. acknowledges support from CONAHCyT grant number 289714 and funding from the National Science and Technology Council (NSTC 112-2636-M-003-001).
M.Z.A. acknowledges support from CONAHCYT grant number {\tt 320772}.
J.B.P.  acknowledges UNAM-DGAPA-PAPIIT support through grant number {\tt IN-111-219}, CONAHCYT, through grant number {\tt 86372}, and to the Paris-Saclay University’s Institute Pascal for the invitation to ‘The Self-Organized Star Formation Process’ meeting, in which invaluable discussions with the participants lead to the development of the idea behind this work. 
A.B.B. acknowledges scholarship by CONAHCyT. 

\section*{Data Availability}

The data underlying this article will be shared on reasonable request to the corresponding author.



\bibliographystyle{mnras}
\bibliography{biblio,other_refs} 






\bsp	
\label{lastpage}
\end{document}

%% file: definitions.tex
\def\apj{ApJ}
\def\apjl{ApJL}
\def\aap{A\& A}
\def\nat{Nature}
\def\araa{ARA\&A}
\def\mnras{MNRAS}

\def\apjs{ApJS}

\def\mnras{{MNRAS}}

\def\alamenos#1{$^{-#1}$}

\newcommand{\kms}{\rm {km~s^{-1}}}

\newcommand{\Msun}{$M_\odot$}    %
\newcommand{\taudynstars}{\tau_{\rm dyn,*}}

\newcommand{\tauff}{\tau_{\rm ff}}

\newcommand{\taurelax}{\tau_{\rm relax}}
\newcommand{\taurelaxstars}{\tau_{\rm relax,*}}